\documentclass[preprint,aps,12pt,showpacs,tightenlines,nofootinbib]{revtex4}
\usepackage{amsmath}
\usepackage{amssymb}
\usepackage{epsfig}
\usepackage{graphicx}
\begin{document}
\preprint{NJNU-TH-07-01}
\newcommand{\beq}{\begin{eqnarray}}
\newcommand{\eeq}{\end{eqnarray}}
\newcommand{\non}{\nonumber\\ }
\newcommand{\mw}{M_W }
\newcommand{\calm}{ {\cal M}}
\newcommand{\ov}{ \overline }


\newcommand{\psl}{ p \hspace{-1.8truemm}/ }
\newcommand{\ksl}{ k \hspace{-1.8truemm}/ }
\newcommand{\qsl}{ q \hspace{-1.8truemm}/ }

\def \cpl{ Chin. Phys. Lett.  }
\def \ctp{ Commun. Theor. Phys.  }
\def \epjc{ Eur. Phys. J. C }
\def \jpg{  J. Phys. G }
\def \npb{  Nucl. Phys. B }
\def \plb{  Phys. Lett. B }
\def \prd{  Phys. Rev. D }
\def \prl{  Phys. Rev. Lett.  }

\title{Exact calculations of vertex $\bar{s}\gamma b$ and $\bar{s} Z b$ in the unitary gauge}
\author{Lu-sen Wu}
\author{Zhen-jun Xiao} \email{xiaozhenjun@njnu.edu.cn}
\affiliation{Department of Physics and Institute of Theoretical Physics, Nanjing Normal
University, Nanjing, Jiangsu 210097, P.R.China}
\date{\today}
\begin{abstract}
In this paper, we present the exact calculations for the vertex $\bar{s}\gamma b$ and $\bar{s} Z b$
in the unitary gauge. We found that (a) the divergent- and $\mu$-dependent terms are left
in the effective vertex function $\Gamma^\gamma_\mu(p,k)$ for $b \to s \gamma$ transition
even after we sum up the contributions from four related Feynman diagrams;
(b) for an on-shell photon, such terms do not contribute et al;
(c) for off-shell photon, these terms  will be canceled when the contributions from
both vertex $\bar{s}\gamma b$ and $\bar{s} Z b$ are taken into account simultaneously, and
therefore the finite and gauge independent function $Z_0(x_t)=C_0(x_t)+ D_0(x_t)/4$, which governs
the semi-leptonic decay $b \to s l^- l^+$, is derived in the unitary gauge.
\end{abstract}


\maketitle

\section{Introduction}

The  inclusive rare radiative decays of b quark, $b \to s V$ with $V=\gamma, Z$ followed by the decays
$V \to q\bar{q}, l^+ l^-$, are induced by the flavor changing neutral current (FCNC) and
play an important role in testing the standard model (SM) and searching for the signal or
evidence of the new physics beyond the SM \cite{hurth03}.

Because of its great importance, these rare decays have been studied extensively during the
past three decades.
The $s \to d V$ transitions with $V=\gamma, Z$ were firstly calculated in Ref.~\cite{inami81} in
$R_\xi$ gauge.
The general transition $q_i \to q_j  \gamma$ for arbitrary quark flavors with arbitrary mass
of internal quarks
were evaluated in the 't Hooft-Feynman gauge \cite{desh-2463} or in a non-linear $R_\xi$ gauge \cite{desh83},
where the current conservation was used to renormalize the vertex $\bar{q}_j \gamma q_i$.

In Ref.~\cite{chia1}, an exact calculation of the $\bar{d}sg$ vertex was made in the unitary gauge, where
the Ward-Takahashi identity was used to renormalize the vertex.
For the quark-photon vertex $\bar{d}s\gamma$ in the unitary gauge, it was firstly
calculated in Ref.~\cite{chia2}, but the results given there were not correct.
In Ref.~\cite{chia3},
the author studied the radiative decay $ b \to s \gamma $ directly and presented their new
results for the on-shell vertex function $\Gamma_\mu^\gamma(p,k)$. The analytical results presented in
Refs.~\cite{inami81,desh-2463,desh83} can also be extended easily
to the case of vertex $\bar{s} \gamma b$, the relevant
basic functions $D_0(x_t)$ and $D_0^\prime(x_t)$ \cite{buras91,buras96} can be extracted directly.
Of course, it is worth of mentioning that the basic function $D_0(x_t)$, as well as
$B_0(x_t)$ and $C_0(x_t)$ are all gauge dependent \cite{buras91,buras96}.

In this paper, we present the results of recalculation of  the $\bar{s}\gamma b$ vertex
and the first calculation of $\bar{s} Z b$ vertex in the unitary gauge.
This paper is organized as follows. In Sec.~\ref{sec:sec2}, we recalculate the $\bar{s}\gamma b$ vertex
in the unitary gauge and extract out  the basic function $D_0(x_t)$ and $D_0^\prime(x_t)$.
In Sec.~\ref{sec:sec3}, we calculate the $\bar{s}Z b$ vertex in the unitary gauge,
derive the basic function $C_0(x_t)$ and make a linear combination of $Z_0(x_t)=C_0(x_t) + D_0(x_t)/4$.
The summary and some discussions are also included in Sec.~\ref{sec:sec3}.

\section{The effective $\bar{s}\gamma b$ coupling } \label{sec:sec2}

\begin{figure}[tb]
\vspace{-3cm}
\centerline{\mbox{\epsfxsize=18cm\epsffile{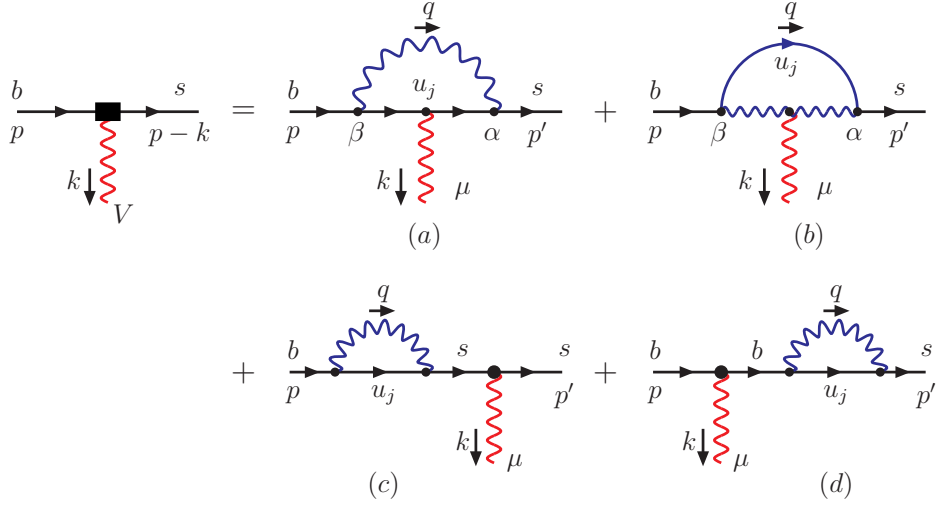}}}
\vspace{-14cm}
\caption{The One-loop Feynman diagrams contributing to the induced $\bar{s} \gamma b$ vertex or
$\bar{s} Z b$ vertex in the unitary gauge.}
\label{fig:fig1}
\end{figure}

In the unitary gauge, the one loop Feynman diagrams for the flavor changing vertex
$\Gamma_\mu^{(\gamma)}(p,k)$ are illustrated in Fig.~\ref{fig:fig1} for the case of $V=\gamma$.

For Fig.~\ref{fig:fig1}a, with the aid of the unitarity relation $\sum_{j=u,c,t}V_{js}^{*}V_{jb}=0$,
 the vertex $\Gamma_{\mu}^{a}(p,k)$ can be written as
\beq
\label{2}
\Gamma_{\mu}^{a}(p,k)&=&\frac{2}{3}e\frac{g^{2}}{32\pi^{2}}\sum_{j}V_{js}^{*}V_{jb} \cdot
 \left \{ \left [ -\frac{3}{4-N} +\frac{3}{2}\ln \left ( \frac{e^{\gamma}M_{W}^{2}}{4\pi\mu^{2}} \right )
-\frac{5}{6} \right ]\; x_j\;  \gamma_{\mu} L + E_{1}\gamma_{\mu}L \right. \non
&&\left. + \left (
E_{2}\psl\gamma_{\mu}\psl+E_{3}\ksl\gamma_{\mu}\ksl+E_{4}\psl\gamma_{\mu}\ksl
    +E_{5}\ksl\gamma_{\mu}\psl+E_{6}\psl\ksl\gamma_{\mu} +E_{7}\gamma_{\mu}\ksl\psl \right )\frac{L}{M_{W}^{2}}
\right. \non && \left.     +E_{8}\psl\ksl\gamma_{\mu}\ksl\psl
\frac{L}{M_{W}^{4}} \right \},
\label{eq:gmu1a}.
\eeq
where $x_j=m^2_j/\mw^2$ with $m_j=(m_u, m_c, m_t)$, $L=(1-\gamma_{5})/2$, $N=4-2\epsilon$ is the dimension
of space-time in dimensional regularization, and the functions
\beq
\label{3}
E_{i}=\int_{0}^{1}dx\int_{0}^{1-x}dy H_{i}, \qquad i=1,2,3,\cdots,8 .
\label{eq:ei8}
\eeq
with
\beq
H_{1} &=& \left \{ 2x_j +\frac{\left [ x(1-x)p^2+y(1-y)k^2-2xyp \cdot k \right ]
\left [ x(1-x)p^2-y^2k^2-2xyp\cdot k \right ]}{\mw^4} \right\} D^{-1}\non
&&  -\left [ 2(1-3x)  -(5-6x) x_j
+ \frac{12x(1-x)p^2 +3y(3-4y) k^2 -24xy p\cdot k }{\mw^2} \right ] \ln D,  \non
  H_{2} &=& \left [2x^2 + (1-x)^2 x_{j} \right ] D^{-1} -\ln D, \non
  H_{3} &=& \left [-2y(1-y) + y^2 x_{j} \right ] D^{-1}, \non
  H_{4} &=& -\left [2x(1-y)+ (1-x)y x_{j} \right ] D^{-1}+ \ln D , \non
  H_{5} &=& \left [2xy -(1-x)y x_{j} \right ]D^{-1}, \non
  H_{6} &=& -(1-x-y)\left\{ \frac{1}{\mw^2}\left [ x(1-x)p^2-y^2k^2-2xyp \cdot k \right ] D^{-1}
  -3\ln D \right\}, \non
  H_{7} &=& y\left\{ \frac{1}{\mw^2} \left [ x(1-x)p^2 + y(1-y)k^2-2xy p \cdot k \right ] D^{-1}
  -3\ln D \right\}, \non
  H_{8} &=& -y(1-x-y) D^{-1},
\label{eq:hi8}
\eeq
with
\beq
D = x +(1-x) x_j - \frac{1}{\mw^2}[x(1-x)p^2+y(1-y)k^2-2xy(p\cdot k)].
\label{eq:dd1}
\eeq
Here the functions $E_i$ and $H_i$ are identical with those presented in Ref.~\cite{chia1}.

When the terms proportional to $p^2/\mw^2, k^2/\mw^2$ and $p\cdot k/\mw^2$ in the functions
$H_i$ and $D$ are neglected, it is easy to do the integration and find the analytical expressions
of function $E_i$. From the analytical results, we confirmed that the vertex function $\Gamma_\mu^a(p,k)$
in Eq.~(\ref{eq:gmu1a}) are identical with those presented in Ref.~\cite{chia1} for
the $\bar{d}sg$ vertex, after making a proper change of the quarks involved and
a replacement of the factor $g_s \lambda^a/2$ to  $ e Q_u$
($Q_u=2/3$ is the charge fraction of the up-type quarks).

For $b \to s g$ decay, Fig.~\ref{fig:fig1}b is irrelevant.
For $b \to s \gamma$ decay, however,  Fig.~\ref{fig:fig1}b must be taken into account.
For  Fig.~\ref{fig:fig1}b, the decay amplitude is of the form
\beq
{\cal M}_{1b} &=& \bar{u}_{s}(p^{\prime})\left\{\int\frac{d^{4}q}{(2\pi)^4}\sum_{j}{\frac{ig}{\sqrt{2}}}
 {\gamma_{\alpha}}LV_{js}^{*}\frac{i}{\psl-\qsl-m_{j}}
         {\frac{ig}{\sqrt{2}}}\gamma_{\beta}L
         V_{jb} \right.\non
        &&\cdot {\frac{i}{(k-q)^2-M_{W}^2}}
  \left [ -g^{\alpha\lambda}+\frac{1}{M_{W}^{2}}{(k^{\alpha}-q^{\alpha})}(k^{\lambda}-q^{\lambda})\right ]\non
&&   \cdot ie\left [
(2q_{\mu}-k_{\mu})g_{\nu\lambda}+(2k_{\nu}-q_{\nu})g_{\lambda\mu}+(-k_{\lambda}-q_{\lambda})g_{\mu\nu}\right
] \non &&\cdot \frac{i}{q^2-M_{W}^2}\left. \left
[-g^{\beta\nu}+\frac{q^{\beta}q^{\nu}}{M_{W}^{2}} \right ]\right\}u_b(p) \epsilon^\mu(k).
\label{eq:m1b}
 \eeq
The corresponding vertex $\Gamma_{\mu}^{b}(p,k)$ can then be written as
 \beq
 \Gamma_{\mu}^{b}(p,k)&=&i e\frac{g^2}{2} \sum_{j}V_{js}^{*}V_{jb}\int\frac{d^{4}q}{(2\pi)^4}
 \frac{N_{\mu}\; L}{[(p-q)^{2}-m_{j}^{2}][(k-q)^2-M_{W}^2][q^2-M_{W}^2]},
\label{eq:gmu1b}
 \eeq
where the numerator $N_\mu$ is of the form
\beq
N_{\mu}&=& \gamma_\alpha (\psl -\qsl)\gamma^\alpha (2 q_\mu -k_\mu)
+ \gamma_\mu (\psl -\qsl)(2\ksl -\qsl )
- (\ksl + \qsl)(\psl -\qsl ) \gamma_\mu  \non
&& +\frac{1}{\mw^2}\left [ (\ksl - \qsl)(\psl -\qsl ) \qsl q_\mu
- (\ksl - \qsl)(\psl -\qsl ) q^2 \gamma_\mu + \cdots \right]
\non
&& + \frac{1}{\mw^4}\; (\ksl - \qsl)(\psl -\qsl )\qsl \left ( k\cdot q k_\mu -k^2 q_\mu \right ).
\label{eq:nmu1}
\eeq

By employing the Feynman parametrization and dimensional regularization, we find the expression of the vertex
$\Gamma_{\mu}^{b}(p,k)$
\beq
 \Gamma_{\mu}^{b}(p,k)&=& e\frac{g^2}{32\pi^2}\sum_{j}V_{js}^{*}V_{jb}
 \left \{\left[\frac{3}{4-N}-\frac{3}{2}\ln\left(\frac{e^{\gamma}M_{W}^2}{4\pi \mu^2}\right)
 \right]x_j\gamma_{\mu}L\right.\non
 & & +\left[\frac{1}{2} \frac{1}{4-N}-\frac{1}{4} \ln\left(\frac{e^{\gamma}M_{W}^2}{4\pi \mu^2}\right)
  +\frac{1}{6}\right] \, x_j\, \left (k^{2}\gamma_{\mu}-\ksl  k_{\mu} \right ) \frac{L}{\mw^2}\non
 & &  +E^{\prime}_{1}\gamma_{\mu}L+\left[E^{\prime}_{2}\psl\gamma_{\mu}\psl
  +E^{\prime}_{3}\ksl \gamma_{\mu}\ksl+E^{\prime}_{4}\psl \gamma_{\mu}\ksl\right.\non
 & & \left.\left.+E^{\prime}_{5}\ksl \gamma_{\mu}\psl+E^{\prime}_{6}\psl\ksl
 \gamma_{\mu}+E^{\prime}_{7}\gamma_{\mu}\ksl\psl\right]\frac{L}{M_{W}^2}\right\},
\label{eq:gmu1b2}
\eeq
where the functions $E_i^\prime$ have been defined as
\beq
 E^{\prime}_{i}=\int_{0}^{1}dx\int_{0}^{1-x}dy H^{\prime}_{i}, \qquad i=1,2,3,\cdots,7.
\label{eq:eip}
\eeq
with
\beq
H^{\prime}_{1} &=&
\left\{\frac{1}{M_W^2}\left[-4x(1-x)p^2 +2(1-x-2y+4xy)p \cdot k-2y(1-2y)k^2\right]\right.\non
 & & \left. +\frac{1}{M_W^4}\left[x^3 (1-x)p^4-\frac{1}{2}x(1-x+2y-2y^2 -3xy+4xy^2)p^2 k^2 \right.\right. \non
 & & \left. \left. -4xy(1-x)(1-y)(p \cdot k)^2 -2x^2 (1-x)(1-2y)p^2 (p \cdot k) \right. \right.\non
 && \left. \left. +(1-x-y-3y^2 +2y^3 +xy+5xy^2 -4xy^3)k^2 (p \cdot k)
 -\frac{1}{2}y(1-y-3y^2 +2y^3)k^4\right] \right. \non
&& \left.  +\frac{1}{M_W^6}\left[\frac{1}{2}x^3 (1-x)p^4 k^2
-\frac{1}{2}x^2(1-x-2y+3xy)p^2 k^2 (p \cdot k)+\frac{1}{2}xy(1-y-xy)p^2 k^4 \right.\right.  \non
 & & \left.\left.-xy(1-x)(1-y)k^2 (p \cdot k)^2 +\frac{1}{2}xy^2 (1-y)k^4 (p \cdot k)
 \right]\right\}{D^{\prime}}^{-1}\non
 && +\left\{6+\frac{1}{M_W^2}\left[x(5-7x)p^2 -2(2-3x-4y+7xy)(p \cdot k)
 +(1+\frac{9}{2}y-7y^2)k^2\right]\right.\non
 && \left.+\frac{1}{M_W^4}\left[-\frac{9}{4}x^2 p^2 k^2 -\frac{1}{2}(1-3x-2y+5xy)k^2 (p \cdot k)
 -\frac{1}{4}y^2 k^4\right]\right\}\ln{D^{\prime}}\non
 &&-9D^{\prime}\ln{D^{\prime}}-\frac{3}{4}\frac{k^2}{M_W^2}D^{\prime}\ln{D^{\prime}},
\eeq
\beq
  H^{\prime}_{2} &=& \left\{-2x(1-x) \right.\non
&& \left. - \left (\frac{1}{\mw^2}-\frac{k^2}{2\mw^4} \right )
  \left [x^3(1-x)p^2 -2x^3 y (p \cdot k)+ xy(1-y-xy)k^2 \right ] \right\}{D^{\prime}}^{-1}\non
  &&+x(1+4x)\left(1-\frac{1}{2}\frac{k^2}{M_W^2}\right)\ln{D^{\prime}},
\eeq
\beq
  H^{\prime}_{3} &=& \left\{-y(1-2y)+\frac{1}{2}\frac{1}{M_W^2}
  \left [x(1-x+4y-3xy-4y^2 +2xy^2)p^2\right. \right. \non
  &&\left. -2(1-x)(1-y-3y^2 +2y^3)(p \cdot k)+y(1-y-3y^2 +2y^3)k^2 \right ]\non
  &&\left.+\frac{1}{2}\frac{p \cdot k}{M_W^4}\left [x^2 (1-x-2y+xy)p^2
  +2xy(1-x-y+xy)(p \cdot k)-xy^2 (1-y)k^2 \right ]\right\}{D^{\prime}}^{-1}\non
  && +\left\{\left (-1-\frac{9}{2}y+4y^2 \right )
  +\frac{1}{2}\frac{1}{M_W^2} \left [-x \left (1-\frac{1}{2}x \right )p^2 \right. \right.\non
  &&\left. \left.+(1-3x-2y+5xy)(p \cdot k)+\frac{1}{2}y^2 k^2 \right ] \right\}\ln{D^{\prime}}
  +\frac{3}{4}D^{\prime} \ln{D^{\prime}},
\eeq
\beq
  H^{\prime}_{4} &=& \left\{(1-x-2y+2xy)+\frac{1}{2}\frac{1}{M_W^2}\left [x^2 (1-x-2y+2xy)p^2
  -2x^2 y(1-2y)(p \cdot k)\right.\right. \non
  && \left.\left.  +(1-x-y-3y^2-xy^2+2y^3+2xy^3 ) k^2 \right ]\right.\non
&& \left.    +\frac{1}{2}\frac{p \cdot k}{M_W^4} \left [-x^3 (1-x)p^2
 + 2x^3 y( p \cdot k)-xy(1-y-xy)k^2 \right ]\right\}{D^{\prime}}^{-1}\non
  &&+\frac{1}{2}\left\{(-1-3x+2y+8xy)+\frac{1}{M_W^2}\left [(x+4x^2)(p \cdot k)
  -\frac{k^2}{2}(1-2y) \right ]\right\}\ln{D^{\prime}},
\eeq
\beq
  H^{\prime}_{5} &=& \left\{2xy+\left (\frac{1}{M_W^2}-\frac{1}{2}\frac{k^2}{M_W^4}\right )
  \left [x^2 (1-x-2y+xy)p^2 \right.\right.\non
  && \left. \left.+2xy(1-x)(1-y)(p \cdot k)- xy^2 (1-y)k^2 \right ]\right\}{D^{\prime}}^{-1}\non
  &&+(-3x+4xy)\left(1-\frac{1}{2}\frac{k^2}{M_W^2}\right)\ln{D^{\prime}}, \non
   H^{\prime}_{6} &=& \left\{(2-2x-y)+\frac{1}{2}\frac{1}{M_W^2}[x^2(1-x)p^2 -2x^2 y(p \cdot k)\right.\non
  &&+(-1+x+y-y^2 -xy^2)k^2]+\frac{1}{2}\frac{1}{M_W^4}[x^2 (1-x-2y+xy)p^2 k^2 +2x^3 y(p \cdot k)^2\non
  &&\left.-x^3 (1-x)p^2 (p \cdot k) +xy(1-2x-y+3xy)k^2 (p \cdot k)-xy^2 (1-y)k^4]\right\}{D^{\prime}}^{-1}\non
  &&+\left\{\frac{1}{2}(1-3x-2y)+\frac{1}{M_W^2}\left[\frac{1}{2}x(1+4x)(p \cdot k)
  +\frac{1}{4}(1-6x+8xy)k^2\right]\right\}\ln{D^{\prime}}, \non
  H^{\prime}_{7} &=& \left\{(-2+2x+y)-\frac{1}{M_W^2}\left[x^2 yp^2 -2xy(1-y)(p \cdot k)-y^2 (2-y)k^2\right]
  \right\}{D^{\prime}}^{-1}\non
  &&+\left[(1-2y)-\frac{1}{2}y\frac{k^2}{M_W^2}\right]\ln{D^{\prime}},
\eeq
and
\beq
D^{\prime}=1-x+x x_{j}-\frac{1}{M_{W}^2}[x(1-x)p^{2}+y(1-y)k^{2}-2xyp\cdot k].
\label{eq:dp}
\eeq

For Fig.~\ref{fig:fig1}c and \ref{fig:fig1}d, we find the vertex $\Gamma_\mu^{(c+d)}(p,k)$ in the same way
\beq
\Gamma_\mu^{(c+d)}(p,k)&=&  e \frac{g^2}{32 \pi^2}
\sum_{j}V_{js}^{*}V_{jb}\left\{ \left[ -\frac{1}{4-N}+\frac{1}{2}\ln\left(\frac{e^{\gamma}M_{W}^2}{4\pi \mu^2}
\right)-\frac{2}{9}\right ]x_{j}\gamma_{\mu}L\right. \non
&&+\frac{1}{3}\frac{m_{b}^{2}F(m_{b}^{2})-m_{s}^{2}F(m_{s}^{2})}{m_{b}^{2}-m_{s}^{2}}\gamma_{\mu}L\non
&&\left. +\frac{1}{3}\frac{m_{b}m_{s}(F(m_{b}^{2})-F(m_{s}^{2}))}{m_{b}^{2}-m_{s}^{2}}\gamma^{\mu}R \right\},
\label{eq:gmu1cd}
\eeq
where $(L,R)=(1\mp \gamma_{5})/2$, and  the function $F(p^2)$ is
\beq
  F(p^2)&=&\int_{0}^{1}dx\left[x(2-3x)+(1-x)(4-3x)x_{j}-x(1-x)(5-4x)\frac{p^2}{M_{W}^2}\right]\non
&& \cdot \ln\left[x+(1-x)x_{j}-x(1-x)\frac{p^2}{M_{W}^2}\right].
\label{eq:fp2}
\eeq

By combining the four pieces together, we find the total vertex function $\Gamma_\mu^\gamma(p,k)$
\beq
\Gamma_\mu^\gamma(p,k)&=& \Gamma_\mu^a(p,k)+ \Gamma_\mu^b(p,k)+ \Gamma_\mu^{(c+d)}(p,k)\non
&=& e\frac{g^2}{32\pi^2}\sum_{j}V_{js}^{*}V_{jb} \;
\left[\frac{1}{2} \frac{1}{4-N}-\frac{1}{4}
\ln\left(\frac{e^{\gamma}M_{W}^2}{4\pi \mu^2}\right)+\frac{1}{6}\right]x_{j} \;
\left (k^{2}\gamma_{\mu}-\ksl  k_{\mu} \right ) \frac{L}{\mw^2} \non
 &&+ e\frac{g^2}{32\pi^2}\sum_{j}V_{js}^{*}V_{jb} \; \left \{
 -\frac{7}{9}x_{j}\gamma_{\mu}L
 +\frac{1}{3}\frac{m_{b}^{2}F( m_{b}^{2})-m_{s}^{2}F(m_{s}^{2})}{m_{b}^{2}-m_{s}^{2}}\gamma_{\mu}L\right.\non
&&\left.  +\frac{1}{3}\frac{m_{b}m_{s}\left ( F(m_{b}^{2})-F(m_{s}^{2}) \right )}{
m_{b}^{2}-m_{s}^{2}}\gamma_{\mu}R  +\widetilde{E}_{1}\gamma_{\mu}\; L \right. \non
&&\left. +\left(\widetilde{E}_{2}\psl\gamma_{\mu}\psl+\widetilde{E}_{3}\ksl
\gamma_{\mu}\ksl  +\widetilde{E}_{4}\psl\gamma_{\mu}\ksl
+\widetilde{E}_{5}\ksl\gamma_{\mu}\psl+\widetilde{E}_{6}\psl\ksl\gamma_{\mu}
+\widetilde{E}_{7}\gamma_{\mu}\ksl\psl \right)\frac{L}{M_{W}^{2}}\right.\non
&& \left.  +\widetilde{E}_{8}\psl\ksl\gamma_{\mu}\ksl\psl\frac{L}{M_{W}^{4}} \right\}
\label{eq:gtot}
\eeq
where
\beq
\widetilde{E}_{i}&=& \frac{2}{3} E_{i}+ E^{\prime}_{i}, \qquad i=1,2,3,\cdots,7; \\
\widetilde{E}_8&=& \frac{2}{3} E_8.
\label{eq:eitot}
\eeq

Following Ref.~\cite{inami81}, the renormalization of the vertex $\bar{s} \gamma b$ is achieved here
by summing up the contributions from all four Feynman diagrams in Fig.~\ref{fig:fig1}.
For a given process, in principle, the divergent terms from individual Feynman diagrams should be
canceled each other when combining all contributions.
In fact, the divergent term and the $\mu$-dependent term proportional to $x_j \gamma_\mu L$ in
Eqs.~(\ref{eq:gmu1a},\ref{eq:gmu1b2}) and (\ref{eq:gmu1cd}) do cancel each other, but the terms
in the square brackets proportional to $(k^2 \gamma_\mu - \ksl k_\mu)$ in Eq.~(\ref{eq:gmu1b2}),
i.e.,
\beq
T^{div}=\frac{1}{2} \frac{1}{4-N}-\frac{1}{4} \ln\left(\frac{e^{\gamma}M_{W}^2}{4\pi \mu^2}\right)
\label{eq:tdiv}
\eeq
are still left in the total vertex function $\Gamma_\mu^\gamma(p,k)$.
The term $T^{div}$ arises from Fig.~\ref{fig:fig1}b, the terms proportional to
$1/\mw^4$ in Eq.~(\ref{eq:nmu1}).

If we use the Ward-Takahashi identity directly to renormalize the vertex, as being done in
Refs.~\cite{chia1,chia2}, we find the same vertex
function $\Gamma_\mu^\gamma(p,k)$ as in Eq.~(\ref{eq:gtot}).

Based on previous studies \cite{inami81,buras96}, we know that
the  effective $\bar{s} \gamma b$ coupling can be generally written as
\beq
\bar{s}\gamma b &=& i \bar{s} \; \Gamma_\mu^\gamma(p,k) \;b =
 i \sum_{j}V_{js}^{*}V_{jb}\frac{G_F}{\sqrt{2}}\frac{e}{8 \pi^2}\{D_{0}(x_j)\bar{s}
(k^2 \gamma_{\mu}-\ksl k_\mu)(1-\gamma_5)b\non
&&-i D_{0}^{\prime}(x_j)\bar{s}\, \sigma_{\mu \nu}k^{\nu}\,
\left [ m_s (1-\gamma_5) +  m_b (1+\gamma_5)\right ] b\},
\label{eq:bsgsm}.
\eeq
where the basic functions $D_0(x_j)$ and $D_0^\prime(x_j)$  were obtained many years ago \cite{inami81}
\footnote{In Ref.~\cite{inami81}, the terms $F_1(x_j)$ and $F_2(x_j)$ instead of $D_0(x_j)$ and
$D_0^\prime(x_j)$ were used to denote the two form factors.}
and takes the following form in the 't Hooft-Feynman gauge \cite{inami81,buras91}
\beq
D_0(x_t)&=& \frac{-25 x_t^2 + 19 x_t^3}{36(1-x_t)^3}
 - \frac{8-32 x_t + 54 x_t^2 -30 x_t^3 + 3 x_t^4}{18 (1-x_t)^4} \ln[x_t], \label{eq:d01}\\
D_{0}^{\prime}(x_{t})&=& -\frac{8x_{t}^{3}+5x_{t}^{2}-7x_{t}}{12(1-x_{t})^3}
+\frac{x_{t}^2 (2-3x_t)}{2(1-x_{t})^{4}}\ln[x_{t}]
\label{eq:d0p1},
\eeq
where $x_t=m_t^2/\mw^2$.

In order to compare our results with those given in Eqs.~(\ref{eq:d01}) and (\ref{eq:d0p1}), we do the
integration and extract out the corresponding basic function $D_0(x_t)$ and $D_0^\prime(x_t)$ from vertex
function $\Gamma_\mu^\gamma(p,k)$ in Eq.~(\ref{eq:gtot}).
Since $m_b^2, m_s^2, k^2$ are all much smaller than $\mw^2$, the terms proportional to
$p^2/\mw^2, k^2/\mw^2$ and $k\cdot p/\mw^2$ in functions $H_i$ and $H_i^\prime$
can be neglected safely. The $x$- and $y$-integrations
in Eq.(\ref{eq:eitot}) can therefore be evaluated easily.
We then find the analytical expressions for the
functions $D_0(x_t)$ and $D_0^\prime(x_t)$ in the unitary gauge:
\beq
D_0(x_t)&=& T^{div}  + \frac{153x_t - 383 x_t^2 + 245 x_t^3 - 27 x_t^4}{72 (1 - x_t)^3}\non
&& - \frac{16-64 x_t + 36 x_t^2 +93 x_t^3 - 84 x_t^4 + 9 x_t^5}{36(1-x_t)^4} \ln[x_t], \label{eq:d01b}\\
D_{0}^{\prime}(x_{t})&=& -\frac{8x_{t}^{3}+5x_{t}^{2}-7x_{t}}{12(1-x_{t})^3}
+\frac{x_{t}^2 (2-3x_t)}{2(1-x_{t})^{4}}\ln[x_{t}]
\label{eq:d0p1b},
\eeq
where $x_t=m_t^2/\mw^2$ and the term $T^{div}$ has been defined in Eq.~(\ref{eq:tdiv}).
It is easy to see that
\begin{itemize}
\item
The basic function $D_0^\prime(x_t)$ in both 't
Hooft-Feynman and unitary gauge are identical, since $D_0^\prime(x_t)$ itself is gauge
independent\cite{buras91,buras96};

\item
The basic function $D_0(x_t)$ in Eq.~(\ref{eq:d01b}) is different from that in the 't Hooft-Feynman gauge.
The reason is that the basic function $D_0(x_t)$, as well as $B_0(x_t)$ and $C_0(x_t)$
are all gauge-dependent \cite{buras91}. The basic function $B_0(x_t)$ and $C_0(x_t)$ come from the
box diagrams of $B^0-\bar{B}^0$ mixing and from the $Z-$penguin diagrams with internal top quark
propagators, respectively.

\end{itemize}

For the case of a real photon emission, the photon is on-shell which means $k^2=0$, and
the $k_\mu$ term can not contribute either by using the transverse polarization condition
$k\cdot \epsilon =0$. For a on-shell $\gamma$, therefore, only the term proportional to
$D_0^\prime(x_t)$ in Eq.~(\ref{eq:bsgsm}) contribute, the physics is consequently independent
of the gauge.
For a time-like photon ($k^2>0$), the divergent term $T^{div}$ may induce a serious problem!
But for an off-shell $\gamma$, the photon
will interact with other particles, new Feynman diagrams will be involved and need to be sum up,
which should lead to a finite and gauge independent final result.

For the inclusive $b\to s l^- l^+$ decays, for example, the contributions from both
decay chains $b \to s \gamma^* \to s l^- l^+$  and $b \to s Z^* \to s l^- l^+$ should  be considered
simultaneously. We indeed find that the term $T^{div}$ in Eq.~(\ref{eq:tdiv}) is canceled
by the corresponding term in the vertex $\Gamma_\mu^Z(p,k)$ when the contribution from the
decay chain $b \to s Z^* \to s l^-l^+$ was also taken into account.
We will show this point explicitly in next section.

As mentioned in  the introduction, the Fig.~\ref{fig:fig1}b was firstly calculated in the unitary
gauge in Ref.~\cite{chia2}, but the analytical results given there were wrong.
In Ref.~\cite{chia3}, the author studied the radiative decay $ b \to s \gamma $ directly
in the unitary gauge and presented their new results for the on-shell vertex function
$\Gamma_\mu^\gamma(p,k)$:
\beq
\Gamma_\mu^{\gamma}(p,k)&=& \frac{e G_F}{4 \sqrt{2}\pi^2} \sum_j V_{js}^* V_{jb}
\left [ A_{tot}(x_j) \left ( k_\mu \ksl - k^2 \gamma_\mu\right )\, L \right. \non
&& \left. + i B_{tot}(x_j)\; \sigma_{\mu\nu} k^\nu \left ( m_s L + m_b R \right ) \right], \label{eq:ga-240}
\eeq
where $L, R=(1\mp \gamma_5)/2$, and
\beq
A_{tot}(x_t)&=& -D_0(x_t), \label{eq:atot}\\
B_{tot}(x_t)&=& D_0^\prime(x_t), \label{eq:btot}
\eeq
by definition. Using the formulae as given in Eqs.(6-13) of ref.~\cite{chia3}, one finds the analytical
expressions of $D_0(x_t)$ and $D_0^\prime(x_t)$
\beq
D_0(x_t)&=&-A_{tot}(x_t)= \frac{63 x_t - 151 x_t^2  + 82 x_t^3}{ 36 (1 - x_t)^3}\non
&& - \frac{16 - 127 x_t + 234 x_t^2 - 123 x_t^3 + 6 x_t^4}{36 (1- x_t)^4}, \label{eq:atot2}\\
D_0^\prime(x_t)&=&B_{tot}(x_t)= -\frac{8x_{t}^{3}+5x_{t}^{2}-7x_{t}}{12(1-x_{t})^3}
+\frac{x_{t}^2 (2-3x_t)}{2(1-x_{t})^{4}}\ln[x_{t}]. \label{eq:btot2}
\eeq
It is easy to see that
\begin{itemize}
\item
The gauge independent function $D_0^\prime(x_t)$ as given in
Eqs.(\ref{eq:d0p1},\ref{eq:d0p1b}) and (\ref{eq:btot2}) are identical.

\item
For function $D_0(x_t)$, however, the two expressions in
Eqs.(\ref{eq:d01b}) and (\ref{eq:atot2}) are different.

\end{itemize}

In Ref.~\cite{buras91}, the gauge independent function $Z_0(x_t)$, which governs the semi-leptonic
decay $b \to s l^+ l^-$, was defined as a linear combination of
$C_0(x_t,\xi)$ and $ D_0(x_t,\xi)$,
\beq
Z_0(x_t) &=& C_0(x_t,\xi) + \frac{1}{4} D_0(x_t,\xi)\non
&=& \frac{108x_{t}-259 x_t^2 + 163 x_t^3 - 18x_{t}^{4}}{144(1- x_{t})^3}
    -\frac{8-50 x_t + 63 x_t^2 + 6 x_t^3 -24 x_{t}^{4}}{72(1-x_{t})^4}\ln[x_{t}],
\label{eq:z0xt}
\eeq
where
\beq
C_0(x_t, \xi) = C_0(x_t) + \frac{1}{2} \bar{\rho}(x,\xi), \quad
D_0(x_t, \xi) = D_0(x_t) - 2\bar{\rho}(x,\xi), \label{eq:d0xi}
\eeq
with
\beq
\bar{\rho}(x,\xi)&=& 2 \rho(x_t,\xi) - 7 B_0(x_t), \label{eq:rhobar}\\
\rho(x,\xi)&=& \frac{\xi}{x-\xi} \left ( \frac{3}{4} \frac{1}{x-1}\right ) x \ln[x]
+ \frac{1}{8} \frac{\xi^2}{x-\xi} \left [\left (\frac{5+ \xi}{1-\xi} -\frac{\xi}{x-\xi} \right )\ln[\xi]
-1 \right ] \non
&& - \left ( x\to 0\right), \label{eq:rhoxi}\\
B_0(x_t)&=& \frac{x_t}{4(1-x_t)} + \frac{x_t}{4(1-x_t)^2} \ln[x-t]. \label{eq:b0xt}
\eeq

In next section, we will calculate the $Z$-penguin diagrams, extract out the basic function $C_0(x_t)$, and
check the gauge independence of function $Z_0(x_t)=C_0(x_t)+ D_0(x_t)/4$ in the unitary gauge.

\section{The effective $\bar{s} Z b$ coupling}\label{sec:sec3}

In the unitary gauge, the one loop Feynman diagrams for the flavor changing vertex
$\Gamma_\mu^{Z}(p,k)$ are illustrated in Fig.~\ref{fig:fig1} for the case of $V=Z$ with a
virtual $Z$ ($k^2>0$).

Following the same procedure as in last section, one can find the vertex functions from
Fig.~\ref{fig:fig1}a-1d for the case of $V=Z$.
First, the individual decay amplitudes can be written as
\beq
{\cal M}_Z^{(a)}&=&\bar{u}_s(p^{\prime})
\left\{\int\frac{d^{4}q}{(2 \pi)^4}\sum_{j}\frac{ig}{\sqrt{2}}\gamma_{\alpha}LV_{js}^{*}
\frac{i}{\psl-\qsl-\ksl-m_{j}}\right.\non
&& \cdot \frac{ig}{4\cos\theta}\gamma_{\mu}[(1-\frac{8}{3}\sin^2 \theta)-\gamma_5]
\frac{i}{\psl-\qsl-m_{j}}\non
&&  \cdot  \left.\frac{ig}{\sqrt{2}}\gamma_{\beta}LV_{jb}\frac{i}{q^2-M_{W}^2}(-g^{\alpha\beta}
+\frac{q^{\alpha}q^{\beta}}{M_{W}^{2}})\right\}\;
u_b(p) \epsilon^\mu(k), \label{eq:m2a}
\eeq
\beq
{\cal M}_Z^{(b)}&=&\bar{u}_{s}(p^{\prime})\left\{\int\frac{d^{4}q}{(2\pi)^4}\sum_{j}{\frac{ig}{\sqrt{2}}}
 {\gamma_{\alpha}}LV_{js}^{*}\frac{i}{\psl-\qsl-m_{j}}
         {\frac{ig}{\sqrt{2}}}\gamma_{\beta}L
         V_{jb} \right.\non
        &&\cdot {\frac{i}{(k-q)^2-M_{W}^2}}
  \left [ -g^{\alpha\lambda}+\frac{1}{M_{W}^{2}}{(k^{\alpha}-q^{\alpha})}(k^{\lambda}-q^{\lambda})\right ]\non
&&   \cdot ig \cos \theta \left [
(2q_{\mu}-k_{\mu})g_{\nu\lambda}+(2k_{\nu}-q_{\nu})g_{\lambda\mu}+(-k_{\lambda}-q_{\lambda})g_{\mu\nu}\right
] \non &&\cdot \frac{i}{q^2-M_{W}^2}\left. \left
[-g^{\beta\nu}+\frac{q^{\beta}q^{\nu}}{M_{W}^{2}}  \right ]\right\}u_b(p) \epsilon^\mu(k), \label{eq:m2b}
\eeq
\beq
{\cal M}_{Z}^{(c+d)}&=&  \bar{u}_{s}(p^{\prime})\left\{-\frac{ig}{2 \cos \theta}
\cdot \frac{g^2}{32 \pi^2} \sum_{j}V_{js}^{*}V_{jb}\left[(-1+\frac{2}{3}\sin^2 \theta)
\left(-\frac{3}{4-N}\right.\right.\right. \non
&&\left. +\frac{3}{2}\ln\left(\frac{e^{\gamma}M_{W}^2}{4\pi \mu^2}\right)-\frac{2}{3}\right)x_{j}
\gamma_{\mu}L
+(-1+\frac{2}{3}\sin^2 \theta)\frac{m_{b}^{2}F(m_{b}^{2})-m_{s}^{2}F(m_{s}^{2})}{m_{b}^{2}-m_{s}^{2}}
\gamma_{\mu}L\non
&&\left.\left.+\frac{2}{3}\sin^2 \theta \frac{m_{b}m_{s}(F(m_{b}^{2})-F(m_{s}^{2}))}{m_{b}^{2}-m_{s}^{2}}
\gamma_{\mu}R \right]
 \right\}u_{b}(p)\epsilon^\mu(k).
\label{eq:m2cd}
\eeq

Taking the same approximation as in last section,
the corresponding vertex function $\Gamma_\mu^{Z}(p,k)$ can be derived from
Eqs.~(\ref{eq:m2a})-(\ref{eq:m2cd}) after making the $x$- and $y$-integrations:
\beq
\Gamma_\mu^{(Z,a)}(p,k)&=&\frac{g}{2\cos\theta} \frac{g^2}{32 \pi^2} \sum_{j}V_{js}^{*}V_{jb}
\left\{ \left ( 1-\frac{4}{3}\sin^2\theta \right )
\left [-\frac{3}{4-N}+\frac{3}{2}\ln \left ( \frac{e^{\gamma}M_{W}^{2}}{4\pi\mu^{2}} \right )
-\frac{5}{6} \right ] x_{j} \right.\non
&&\left.  + \left(1-\frac{4}{3}\sin^2\theta \right)\left[\frac{8-31x_j +5{x_{j}}^2}{12(1-x_j)}
-\frac{3{x_j}^{2}(2-x_j)}{2(1-x_j)^2}\ln x_j\right]
\right.\non
&& \left.-\left[\frac{1}{4-N} -\frac{1}{2}\ln \left( \frac{e^{\gamma}M_{W}^{2}}{4\pi\mu^{2}}
\right)-\frac{1}{2}\right]x_j
\right.\non
&&\left. + \frac{x_j (5+x_j)}{4(1-x_j)}+ \frac{x_j ({x_j}^2 -2x_j +4 )}{2(1-x_j)^2}\ln x_j
\right \}\; \gamma_{\mu}L + \cdots,
\label{eq:gmuza}
\eeq
\beq
\Gamma_\mu^{(Z,b)}(p,k)&=& g\cos\theta \frac{g^2}{32 \pi^2}\sum_{j}V_{js}^{*}V_{jb}
\left\{\left[\frac{3}{4-N}-\frac{3}{2}\ln\left(\frac{e^{\gamma}
M_{W}^2}{4\pi \mu^2}\right)\right] x_j \right.\non
&& \left. +\frac{-2+13x_j -5{x_j}^2}{4(1-x_j)}+\frac{3{x_j}^2 (2-x_j)}{2(1-x_j)^2}\ln x_j\right\}
\gamma_{\mu}L  + \cdots, \label{eq:gmuzb}
\eeq
and
\beq
\Gamma_\mu^{(Z,c+d)}(p,k)&=& \frac{g}{2 \cos \theta} \left (1 - \frac{2}{3}\sin^2 \theta \right )
\frac{g^2}{32 \pi^2} \sum_{j}V_{js}^{*}V_{jb}\non
&& \cdot \left\{\left[-\frac{3}{4-N}+\frac{3}{2}
\ln\left(\frac{e^{\gamma}M_{W}^2}{4\pi \mu^2}\right) -\frac{2}{3}\right] x_{j} \right.\non
&& \left. -\frac{2 +23 x_j -7{x_j}^2}{12(1-x_j)}-\frac{3{x_j}^2 (2-x_j)}{2(1-x_j)^2}\ln x_j \right\}
\gamma_{\mu}L  + \cdots . \label{eq:gmuzcd}
\eeq
where only the terms proportional to $\gamma_\mu L$ are shown explicitly. The total
vertex function $\Gamma_\mu^{Z}(p,k)$ can therefore be written as
\beq
\Gamma_\mu^{Z}(p,k) &=& \frac{G_F}{\sqrt{2}}\frac{e}{2 \pi^2}M_{Z}^2
\frac{\cos\theta}{\sin\theta}\sum_{j}V_{js}^{*}V_{jb}
\frac{1}{8}\left\{ \left[-\frac{1}{4-N}+\frac{1}{2}\ln\left(\frac{e^{\gamma}M_{W}^2}{4\pi \mu^2}\right)\right]
x_j\right.\non
&&\left. + \frac{7x_j -8x_{j}^2+x_{j}^3 }{4(1-x_j)^2}
+\frac{4x_j -2x_j^2+x_j^3\ln{x_j}}{2(1-x_j)^2} \right\}\; \gamma_{\mu}(1-\gamma_5) + \cdots,
\label{eq:gmu2}
\eeq
where the relation $e=g \sin{\theta}$, $\mw=M_Z \cos\theta$ and $G_F=g^2/(4\sqrt{2}\mw^2)$ have been
used.

By comparing the vertex function $\Gamma_\mu^{Z}(p,k)$ in Eq.~(\ref{eq:gmu2}) with
the general expression of the effective coupling $\bar{s}Zb$ \cite{buras96},
\beq
\bar{s}Z b=i \sum_{j}V_{js}^{*}V_{jb}\frac{G_F}{\sqrt{2}}\frac{e}{2 \pi^2}M_{Z}^2
\frac{\cos\theta}{\sin\theta}C_{0}(x_j)\bar{s}\gamma_{\mu}(1-\gamma_5)b,
\eeq
one can extract out the basic function $C_0(x_t)$ in the unitary gauge
\beq
 C_{0}(x_{t})&=& \frac{1}{8}\left[-\frac{1}{4-N}+\frac{1}{2} \ln\left(\frac{e^{\gamma}M_{W}^2}{4\pi
 \mu^2}\right)\right] x_t \non
 &&  + \frac{7x_{t}-8x_{t}^{2}+x_{t}^3}{32(1-x_{t})^2}
+ \frac{4x_{t}-2x_{t}^{2}+x_{t}^3}{16 (1-x_t)^2}\ln[x_t], \label{eq:c0xt2}
\eeq
where $x_t=m_t^2/\mw^2$. Again, a divergent- and a $\mu$-dependent term also appear here!

By combing the function $D_0(x_t)$ in Eq.~(\ref{eq:d01b}) with the function $C_0(x_t)$ in
Eq.~(\ref{eq:c0xt2}), we find the function $Z_0(x_t)$ in the unitary gauge
\beq
Z_0(x_t) &=& C_0(x_t) + \frac{1}{4} D_0(x_t) \non
&=& \frac{108x_{t}-259 x_t^2 + 163 x_t^3 - 18x_{t}^{4}}{144(1- x_{t})^3}
    -\frac{8-50 x_t + 63 x_t^2 + 6 x_t^3 -24 x_{t}^{4}}{72(1-x_{t})^4}\ln[x_{t}],\label{eq:z0xt2}
\eeq
which is indeed finite, gauge independent and identical with the one as given in
Eq.~(\ref{eq:z0xt}) \cite{buras91,buras96}. It is worth of stressing that
the divergent- and $\mu$-dependent  terms in both $C_0(x_t)$ and $D_0(x_t)$ in the unitary gauge
was canceled in the linear combination, and the gauge independence of $Z_0(x_t)$ is true in unitary gauge.

\begin{figure}[tb]
\vspace{-4cm}
\centerline{\mbox{\epsfxsize=18cm\epsffile{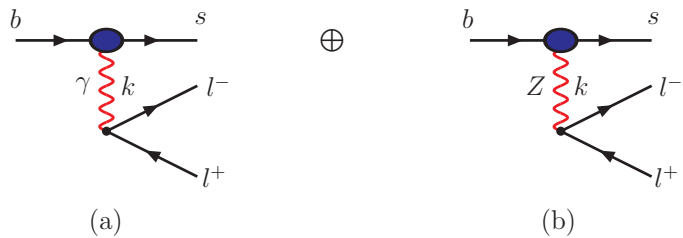}}}
\vspace{-17cm}
\caption{Two ``tree" Feynman diagrams contributing to $b \to s l^- l^+$ decays, where the oval denotes
the $\bar{s} \gamma b$ vertex and $\bar{s} Z b$ vertex ($k^2>0$), respectively.}
\label{fig:fig2}
\end{figure}

For the semi-leptonic decays $b \to s l^+ l^-$, as illustrated in Fig.~\ref{fig:fig2}, it can proceed
through the decay chain $b \to s \gamma^* \to s l^- l^+$ or $b \to s Z^* \to s l^- l^+$, and
therefore  both effective coupling $ \bar{s} \gamma b $ and $\bar{s} Z b$ should contribute simultaneously.
The divergent-term, the $\mu$-dependent term and gauge dependence appeared in both $D_0(x_t)$ and $C_0(x_t)$
are canceled each other when one sum up all contributions together, which is consistent with the
general expectation.

To summarize, we have recalculated at one-loop level the effective coupling
$\bar{s}\gamma b$ and $\bar{s} Z b$ in
the unitary gauge and extracted the relevant basic functions $C_0(x_t)$, $D_0(x_t)$ and $D_0^\prime(x_t)$.
We start from the calculation of Fig.~1a for the case of $V=\gamma$ and confirmed the analytical
results presented in Ref.~\cite{chia1}, but our calculations for Fig.~1b are different from
the corresponding results as given in Ref.~\cite{chia2,chia3}.
We finally calculated the effective coupling $\bar{s} Z b$ and made a linear combination
of the basic function $D_0(x_t)$ and $C_0(x_t)$ as a means of cross checking.
We found that
(a) the function $D_0^\prime(x_t)$ is identical with those known results;
(b) the divergent- and $\mu$-dependent terms appeared in both $C_0(x_t)$ and $D_0(x_t)$ function
in the unitary gauge are canceled each other in a linear combination $Z_0(x_t)=C_0(x_t) + D_0(x_t)/4$;
and (c) the resultant $Z_0(x_t)$ function, which governs the semi-leptonic $b \to s l^- l^+$ decays, is
finite, gauge independent and identical with those as given in Ref.~\cite{buras91,buras96}.

\begin{acknowledgments}

We are very grateful to Zhao-hua Xiong, Wen-juan Zou and Lin-xia L\"u for helpful discussions.
This work is partly supported  by the National Natural Science
Foundation of China under Grant No.10575052, and by the
Specialized Research Fund for the doctoral Program of higher education (SRFDP)
under Grant No.~20050319008.

\end{acknowledgments}

\end{document}